# Experimental Investigation of Thermal Perfomance for Selected Oils for Solar Thermal Energy Storage and Rural Cooking Application

Karidewa Nyeinga[1], Denis Okello[1], Tabu Bernard[1,2], Ole Jorgen Nydal[3]

[1]Department of Physics, Makerere University, Uganda
[2]Department of Physics, Gulu University, Uganda
[3]Department of Energy & Process Engineering, Norwegian University of Science & Technology (NTNU), Norway.

**Abstract**

The thermal performance of locally available oils in Uganda have been experimentally investigated to determine their suitability for heat transfer and solar thermal energy storage. Sunflower oil and palm oil, both local vegetable vegetable oils and Shell thermia B, commonly used in the local industries as a heat transfer fluid were used in the study. The oils were heated in an insulated tank until temperatures close to their smoke points were attained and then heating was stopped; the tank temperature was recorded for 24 hours during the cooling. The second experiement involved charging the oil in a self-circulating system without the aid of a pump. It was observed that the vegetable oils gained heat faster than thermia B. Sunflower oil retained heat for a longer period compared to palm oil and thermia B. The total amount energy stored by sunflower was equally higher than palm oil while thermia B had the least energy stored. The results show that sunflower oil is the most suitable oil for solar thermal storage and heat transfer.

## 1. Introduction

Solar thermal storage systems for rural cooking applications require suitable heat transfer fluids capable of transferring heat to the storage during charging and to the cooker during discharging (Lovseth 1997; Heetkamp 2002). The choice of fluid for heat transfer depends on the storage medium and the choice for heat extraction. There are industrial fluids available for heat transfer in large scale solar thermal plants. However, such fluids are expensive for rural applications and therefore will increase the cost of the system. Air and water have been used as heat transfer media with rock bed storage systems as reported by (Okello et al. 2016; Nyeinga, et al. 2016). However, air has low heat capacity and water may pause high risks since it vaporizes at high temperatures and therefore not suitable for rural cooking application.

Vegetable oils can be used as both heat transfer fluid and heat storage medium for solar cooking applications. The advantage with vegetable oils is that they are not risky even if there are leakages in the system. Mawire et al. (2014) explored the use of sunflower as heat transfer medium; they studied the charging of the oil using high and low flow rates. For cooking applications, the interest is in oils which are capable of attaining temperatures of about $200^oC$. In this paper, thermal properties of selected oils were investigated

## 2. The Experimental set-up and procedure

### 2.1. Oil samples

Local vegetable oils namely refined sunflower oil and refined palm oil were used in this study in addition to thermial B, a mineral oil but readily available in the country. Table 1 shows the thermo-physical properties of the selected oils. It can be observed that the vegetable oils have high densities and specific heat capacities compared to thermia B.

**Tab. 1: Thermo physical properties of the selected oils adapted from Mawire et al. (2014)**

|  | **Density** $(kgm^{-3})$ | **Specific heat capacity** $(Jkg^{-1}K^{-1})$ | **Thermal conductivity** $(Wm^{-1}K^{-1})$ |
|---|---|---|---|
| Refined sunflower oil | $930.62 - 0.65T$ | $2115.00 + 3.13T$ | $0.061 + 0.018e(-T/26.142)$ |
| Refined palm oil | $925.00 - 0.66T$ | $1861$ | $0.1721$ |
| Thermia B | $870.00 - 0.65T$ | $1798.00 + 3.58T$ | $0.118 + 0.018e(-T/168.660)$ |

*3.2. Determination of heat retention capacity*

A cylindrical tank of about 4.5L fitted with a 1.5kVA electrical heater was used to heat 4L of the selected oil samples in turn. Three K-type thermocouples fitted in the tank at a distance of 5cm apart were used to measure the temperature along the tank. The thermocouples were connected to a TC-08 data logger interfaced with computer as shown in figure 1. The heater was connected to a 240V a.c main and the oil in the tank was heated to a temperature close to its smoke point and the heater switched off. The temperature of in the tank was recorded for about 25 hours.

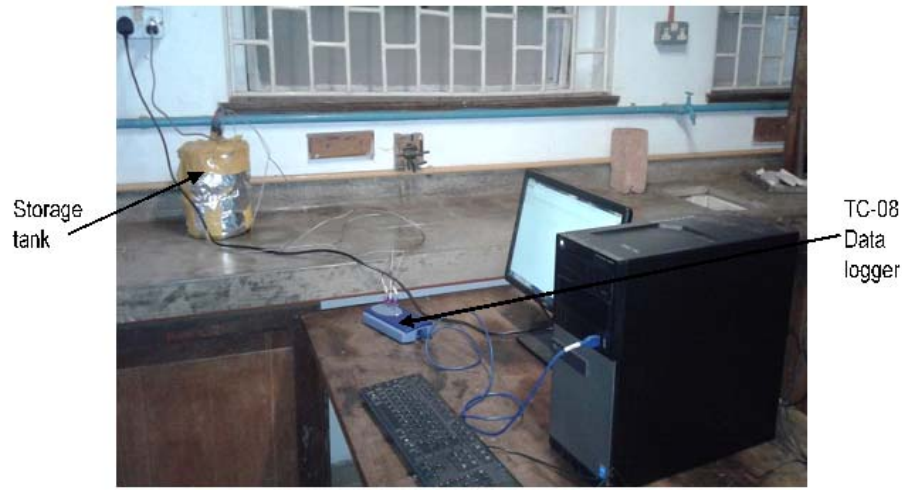

**Fig. 1: Showing the experimental setup. The oil is heated in an insulated tank and thermocouples connected to a data logger and a computer were used to monitor and record the temperature during heating and cooling.**

The quantity of heat energy, $Q$ stored in oil at any time is expressed as:

$$Q = \rho v c (T - T_o) \qquad \text{(eq. 1)}$$

where $\rho$, $c$, $v$ are average density of the oil, average specific heat capacity of the oil and volume respectively. $T$ is the average temperature at time, $t$ and $T_o$ is the initial temperature. The average temperature of the oil in the storage tank was considered in the computation of the energy content of the tank since the storage tank was short.

3.3. Self circulating charging unit

Figure 2 shows a storage tank of internal diameter 18cm and height 40cm made of steel and a boiler of internal diameter 18cm and height 20cm made of mild steel oriented at $60^o$ to the horizontal. The storage tank was charged based on self-circulating system. The main advantage with a self-circulating system is that you avoid the use of pumps. A continuous copper pipe of internal diameter 0.9cm and total length measuring 150cm was inserted from the top of the storage tank to the bottom through the boiler.

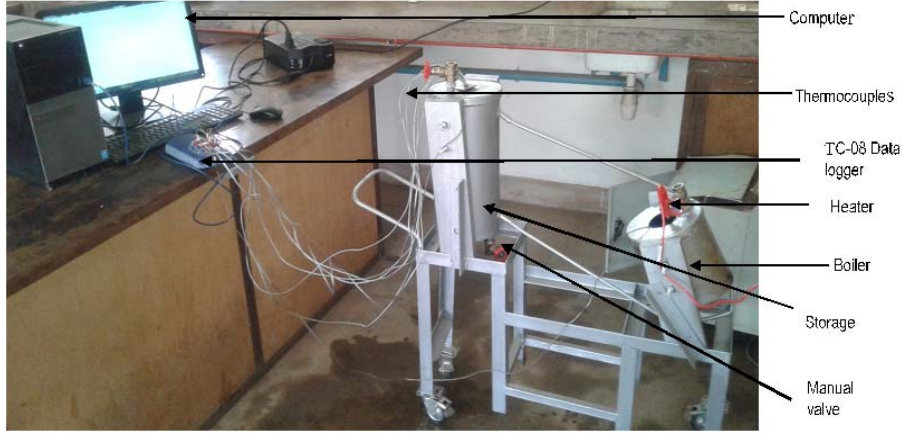

Fig. 2: A self circulating system consisting of a storage tank and a boiler. Sunflower oil was used in the boiler and the three types of oil where put in the storage tank one after the other. The storage was filled with each type of oil and this got heated in the boiler and thereby charging the storage.

Seven k-type thermocouples each at an interval of 5cm were placed at the central axis of the storage tank. The storage tank was filled with 10 litres of sunflower oil followed by palm oil and finally thermia B as a heat storage fluid respectively. The boiler was filled with 4 litres of refined sunflower oil. An electrical heater of 1.5 kVA was used for heating the oil in the boiler. The temperature of the boiler was maintained in the range of $230^oC - 240^oC$ (just below the smoke point of sunflower oil) manually, by switching the electrical heater on and off. The storage tank was charged for 6 hours.

The total energy, $E$ stored in a tank of, $n$ segments during the charging process is given by

$$E = \rho c \sum_{i=1}^{n} v_i \Delta T_i \qquad \text{(eq. 2)}$$

where $\rho$, $c$ are the density of the oil and the specific heat capacity of the oil respectively; $v_i$ is the volume of oil in the $i^{th}$ segment and $T_i$ is temperature difference between two adjacent nodes of the $i^{th}$ segment of a stratified tank.

## 3. Results and discussions
### 4.1 Heat retention
Refined sunflower oil and palm oil were both heated to a maximum temperature of about $236^oC$; these temperatures are slightly below their smoke points. Thermia B was heated to a maximum temperature of $220^oC$. After heating the oils to their maximum temperatures, the heater was switched off and the tank allowed to cool. Figure 4 shows the temperature profiles during heating and cooling. Refined sunflower oil and refined palm oil showed a rapid increase in temperature during the heating while thermia B gained heat slowly.

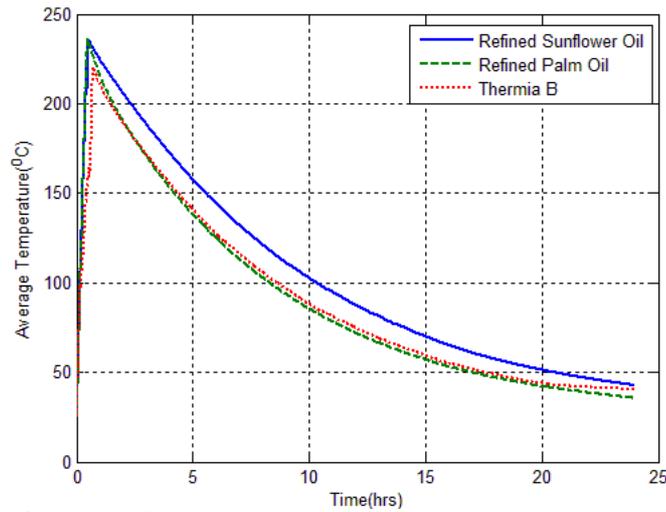

**Fig. 4: Temperature profile of the various oils during heating and cooling with the tank insulated. A rapid increase in temperature of sunflower oil and palm oil are observed during heating. After 24hours, the temperature of sunflower oil can be seen to be higher than for palm oil and thermia B.**

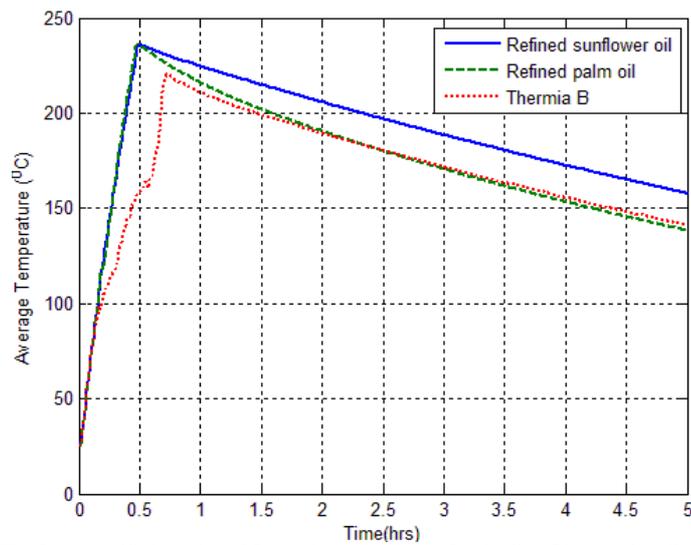

**Fig. 5: a magnified version of figure 4; results shown for the first 5 hours of cooling.**

Both refined sunflower oil and refined palm oil reached a maximum temperature of about $236^oC$ in 30 minutes. This high increase in temperature is attributed to fact that both refined sunflower oil and refined palm oil have high densities and specific heat capacities. However, thermia B took longer to attain its average maximum temperature of $220^oC$ in about 45 minutes because it has low density and specific heat capacity and this can be clearly seen in figure 5 which is an amplified version of in figure 4.

For solar thermal storage systems with cooking applications, we are interested in charging the storage in the shortest possible time since the solar radiation keeps varying with time. Fluids which can retain the heat for a longer period are to be preferred since this allows cooking to be done even after sunset. Both refined sunflower oil and refined palm oil absorbed heat faster than thermia B; but refined refined sunflower retains heat better than palm oil.

*4.2 Energy distribution*

Figure 6 shows the energy distribution during heating and retention computed using equation 1. The maximum energy attained by refined sunflower was 1.6MJ in 30 minutes, while for refined palm oil was 1.4MJ in 30 minutes, and thermia B attained the maximum energy of 1.2MJ in 45 minutes.

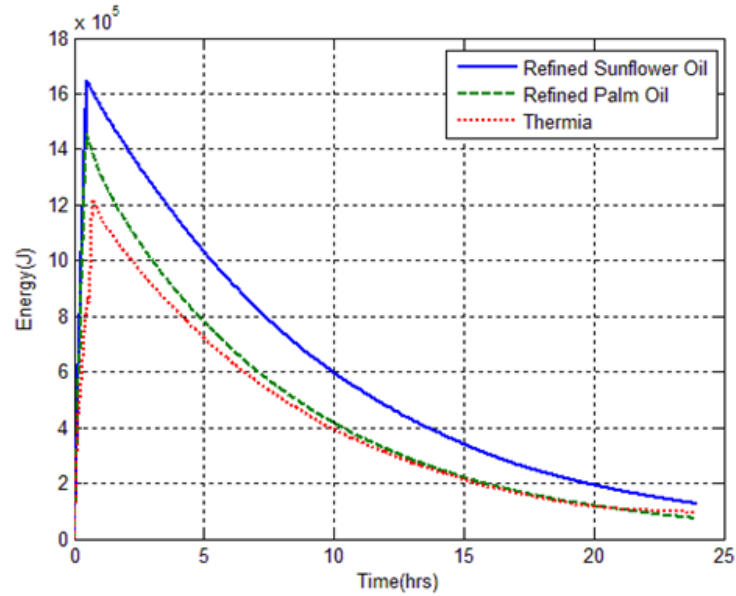

**Fig. 6: energy distribution over time for oils heated for 30-45 minutes and thereafter the heater was switched off and allowed to cool for about 24 hours in an insulated tank.**

From figure 6, it can be observed that at the end of the 24 hours, sunflower oil had much more energy than both palm oil and thermia B. This furthers shows that sunflower oil is more suitable for solar thermal energy applications.

*4.3. Temperature profiles during charging*

Figures 7 shows the temperature profiles in the storage tank based on the self-circulating charging system for 6 hours.

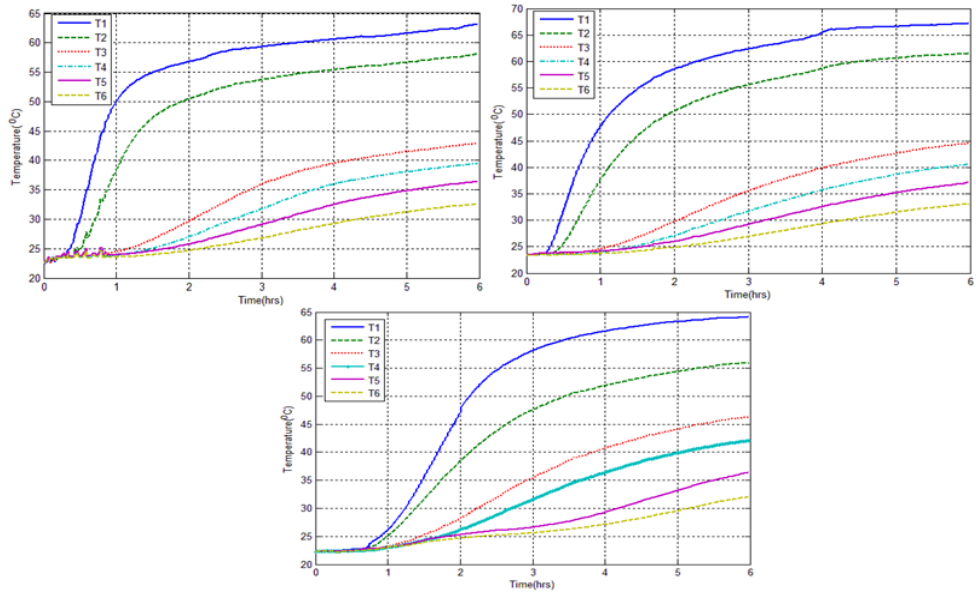

**Fig. 7: Temperature profiles during charging of the storage tank using a self-circulating system for 6 hours using a self-circulating system. $T_1$ is temperature at the top of the storage while $T_6$ is at the bottom. The storage tank was not insulated since the aim was to compare the temperatures during charging of the three different oils under the same conditions.**

In the first one hour, sunflower oil and palm charged very fast and attained temperatures of about $50^oC$ at the top of the tank; however, in the same period thermia B had a temperature below $30^oC$ at the top. The observed high rate of charging attained by sunflower oil and palm oil is attributed to their high densities and specific heat capacities which agrees with similar findings by Mawire et al. (2014).

### 4.4. Thermal energy profiles during charging

Figure 8 shows the energy distribution in the storage tank for the three different oils during charging by a self-circulating system. The energy profiles for both refined sunflower oil and refined palm oil increased rapidly in the first one hour until their peak values were attained. The maximum energy for refined sunflower oil was $0.13MJ$ attained in about 1.5 hours while for refined palm oil was $0.12MJ$ attained in about 2 hours. The energy profile for thermia B increased gradually until it attained a peak value of $0.10MJ$ in about 3 hours. The high energy gained in sunflower in a short time is again associated to its high density and specific heat capacity.

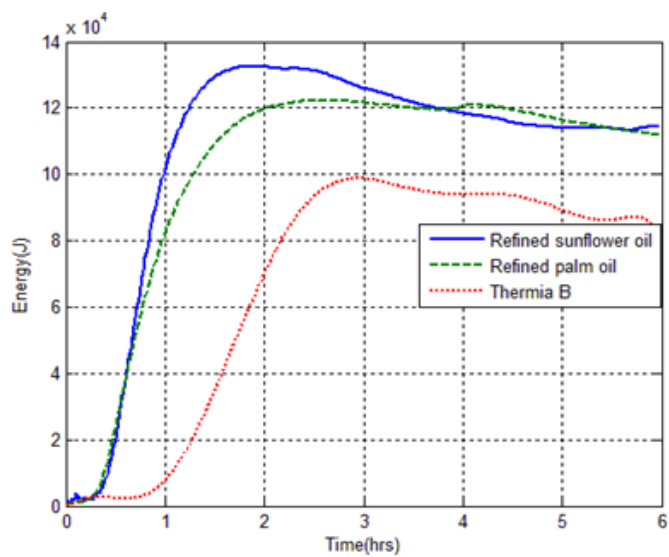

**Fig.8: thermal energy stored in the tank by refined sunflower, refined palm oil and thermia B during charging for 6 hours using a self-circulating system. Sunflower oil the highest energy and thermia B had the least energy.**

The drop in energy after 2.5 hours can be attributed to the fact that the density and specific heat capacity for oil decreases and increases respectively with increasing temperature; but the decrease in density overrides the increase in specific heat capacity as cited by Esteban et al. (2012).

## 5. Conclusion

The thermal performance of refined sunflower oil, refined palm oil, and thermia B was experimentally studied. The results show refined sunflower oil and palm oil gained heat faster than thermia B. However, refined sunflower oil retained more heat than palm oil. For solar energy applications, the solar radiation may be available over a period of time and therefore sunflower oil would be more suitable since it can absorb the heat faster and retain it longer. Further studies should be carried out on a hybrid storage system consisting of rock pebbles and sunflower oil. This will cut down the cost of the storage since rock pebbles are readily available and cheap. We also recommend further work on the number of cycles the sunflower oil can be heated and cooled without any major changes in its properties.

## 6. Acknowledgement

The authors would like to thank NORAD and the Energy & Petroleum (EnPe) Project-Norway, for the financial support received to carry out this research. Further thanks to all project partners for their contribution in this research.